\documentclass[a4paper,11pt]{article}
\usepackage{jheppub}

\pdfoutput=1
\usepackage[T1]{fontenc}
\usepackage{CJK}
\usepackage{xcolor}
\usepackage{amsfonts}
\usepackage{epstopdf}
\usepackage{wrapfig} 
\usepackage{subfigure}
\usepackage{graphicx}  
\usepackage{dcolumn}   
\usepackage{bm}

\newcommand{\be}{\begin{equation}}
\newcommand{\ee}{\end{equation}}
 \newcommand{\bea}{\begin{eqnarray}}
 \newcommand{\ena}{\end{eqnarray}}
  
  \newcommand{\Eq}[1]{Eq.~(\ref{#1})}

\title{Influence of inhomogeneities on holographic mutual information and butterfly effect}

\author[1,2]{{Rong-Gen Cai,}}
\author[1,3]{Xiao-Xiong Zeng,}
\author[4]{Hai-Qing Zhang}

\affiliation[1]{CAS Key Laboratory of Theoretical Physics, Institute of Theoretical Physics,\\ Chinese Academy of Sciences, Beijing, 100190, China}
\affiliation[2]{School of Physical Sciences, University of Chinese Academy of Sciences, Beijing 100049, China}
\affiliation[3]{School of Material Science and Engineering, Chongqing Jiaotong University, Chongqing 400074, China}
\affiliation[4] {Department of Space Science and International Research Institute of Multidisciplinary Science, Beihang University, Beijing 100191, China}

\emailAdd{cairg@itp.ac.cn}
\emailAdd{xxzeng@itp.ac.cn}
\emailAdd{hqzhang@buaa.edu.cn}

\abstract{\\
We study the effect of inhomogeneity, which is induced by the graviton mass in massive gravity, on the mutual information and the chaotic behavior of a 2+1-dimensional field theory from the gauge/gravity duality. When the system is near-homogeneous, the mutual information increases as the graviton mass grows. However, when the system is far from homogeneity, the mutual information decreases as the graviton mass increases. By adding the perturbations of energy into the system, we investigate the dynamical mutual information in the shock wave geometry. We find that the greater perturbations disrupt the mutual information more rapidly, which resembles the butterfly effect in chaos theory. Besides, the greater inhomogeneity reduces the dynamical mutual information more quickly just as in the static case.}

\begin{document}
\maketitle
\flushbottom

\section{Introduction}
\label{sec:intro}
Over the past two decades, there have been a large overlap studies among high energy theory, condensed matter physics, quantum information and geometry based on the conjecture of gauge/gravity correspondence \cite{Maldacena:1997re,Gubser:1998bc,Witten:1998qj}. The typical example is the holographic entanglement entropy (HEE) proposed by Ryu and Takayanagi that relates the minimized codimension-2 bulk surface to the quantum information in the boundary field theory \cite{Ryu:2006bv}. The Ryu-Takayanagi conjecture greatly simplifies the computations of the entanglement entropy in the field theory, and  numerous papers investigates the relation between the bulk geometry and the quantum information in the boundary field theory.  For recent reviews, please refer to Refs.~\cite{VanRaamsdonk:2016exw,Rangamani:2016dms}. Among these studies, one interesting thing is to adopt the HEE to probe the thermalization process or phase transitions in the strongly coupled systems~\cite{Balasubramanian1,Cai:2012sk,Cai:2012nm,Johnson:2013dka,Bai:2014tla,Ling:2015dma,Zeng:2015tfj,Gb,Gb2,Zeng:2014xpa, Zeng:2016sei}.

Mutual information measures  the correlations between two subsystems in quantum information theory \cite{quantuminfor}. The holographic setup of mutual information can be obtained by calculating the length of the wormhole which connects the two sides of an eternal AdS black hole \cite{Maldacena:2001kr,Maldacena:2013xja,VanRaamsdonk:2009ar}. In \cite{Maldacena:2013xja}, the authors defined the thermofield double states (TFD) to describe the states of the entangled two sides of the black hole, i.e.,
\be
|\Psi\rangle\equiv\sum_i e^{-\frac{\beta}{2} E_i}|i\rangle_L\otimes|i\rangle_R,
\ee
where $\beta$ is the inverse of the temperature while $|i\rangle_L$ and $|i\rangle_R$ are the identical quantum states on the two-sided AdS black holes. Supposing there are two identical space-like subregions $A$ and $B$ on each side of the black hole, the mutual information $I(A,B)$ between $A$ and $B$ can be computed as
\be\label{mutual}
I(A,B)\equiv S(A)+S(B)-S(A\cup B),
\ee
where $S(A)$, $S(B)$ are the entanglement entropy of $A$ and $B$ respectively, while $S(A\cup B)$ is the entanglement entropy of the union of $A$ and $B$. Holographically, $S(A)$ and $S(B)$ can be calculated by the areas of the minimal surfaces in the bulk geometry associated to $A$ and $B$ independently from the Ryu-Takayanagi conjecture, while $S(A\cup B)$  can be computed from the minimal surface which crosses the horizon and stretches through the wormhole connecting the two subregions $A$ and $B$ \cite{Morrison:2012iz}.

The disruption of mutual information was related to the {\it butterfly effect} by Shenker and Stanford \cite{Shenker:2013pqa}. Specifically, as a  small perturbation, such as a light-like energy perturbation,  is  added to one side of the eternal black hole, the mutual information between the two sides may be disrupted after an amount of time $t_*$, which means there is no dependence or entanglement between the two sides of the black hole. The time $t_*$, usually called the scrambling time in black hole systems, is proportional to the logarithm of the entropy of the black hole, e.g., $t_*\sim \beta\log(S)/2\pi$, in which $S$ is the Bekenstein-Hawking entropy. The disruption of the mutual information is a piece of evidence of the system's high sensitivity to the initial conditions, which reminds us of the terminology in chaos theory-butterfly effect. Recent  literatures relevant to  the butterfly effect can be found in \cite{Leichenauer:2014nxa,Berenstein:2015yxu,Sircar:2016old,Ling:2016ibq,Qaemmaqami:2017bdn,Alishahiha}.

In this paper, we intend to study the static and dynamical mutual information in the background of massive gravity \cite{deRham:2010ik,deRham:2010kj}. As we know from \cite{Vegh:2013sk,Davison:2013jba,Blake:2013owa} that the graviton mass in the bulk breaks the diffeomorphism invariance, which makes the stress-energy tensor non-conserved in the dual
field theory. The non-conservation of the stress-energy tensor causes the momentum dissipation in the boundary field theory. Therefore, from this aspect the graviton mass plays the role of inhomogeneity in the boundary field theory, i.e., greater graviton mass is dual to greater inhomogeneity in the boundary~\footnote{Strictly speaking, the term ``inhomogeneity'' used here represents the meaning of translational symmetry breaking or momentum dissipation, which was used previously in the framework of holography in the paper \cite{Blake:2013owa}. To be consistent with existing literatures, here we still take the term ``inhomogeneity''  to represent the same meaning. }. There have been a number of papers investigating the inhomogeneous effects caused by massive gravity so far, see for example \cite{Cai:2014znn,Hu:2015xva,Cai:2015wfa,Cao:2015cza,Hu:2015dnl,Hu:2016mym,Zeng:2015tffj,hu20173,Cai:2015cya}.

Firstly, we study the holographic mutual information of two identical strips in a static background of a 3+1-dimensional massive gravity. For the strips with larger length, the mutual information between them decreases monotonically as the graviton mass increases; However, the mutual information between two shorter strips first increases and then decreases with respect to the graviton mass. A plausible reason is that the graviton mass (or equivalently spatial inhomogeneity) would have greater effects on the strips with larger length. When the system is near-homogeneous (or equivalently with small graviton mass), the mutual information for the strips with shorter length will increase with respect to the temperature of the black hole (referring to the right panel of Fig.\ref{fig7} and discussions in the main context), which is similar to the relation between mutual information and temperature in the pure homogeneous case \cite{Leichenauer:2014nxa}. However, as the graviton mass grows big enough, i.e., far from homogeneity, the mutual information will instead decrease with respect to the graviton mass, which behaves distinctly from the homogenous case. Therefore, we argue that the greater graviton mass (or stronger inhomogeneous effects) would play a dominant role in decreasing the mutual information compared with the temperature. On the other hand, if the strips are longer, the mutual information decreases monotonically according to the graviton mass regardless of the system being near or far from homogeneity.  It intuitionaly  suggests that the spatial inhomogeneity would have greater impacts to the strips with larger length by destroying the mutual information monotonically. A detailed discussion of the relation between mutual information and the graviton mass in the static background case  is given in Section \ref{sect:holoMG}.

In all the parameter regimes we considered in this paper, we find that the critical width of the strip, which renders the mutual information vanishing, always decreases according to the graviton mass. Moreover, as the width of the boundary strip increases, it is found that the critical charge which disrupts the mutual information increases as well. As we know, in the massive gravity the temperature of the black hole will decrease as the charge increases. Therefore, we can infer that the critical width of the strip increases when the temperature of the black decreases, which is consistent with the results  obtained in \cite{Leichenauer:2014nxa}.

In order to study the dynamical behavior of mutual information in the boundary field theory, one of the approaches is to add extra energy perturbations into the bulk, which will lead to a shift on the horizon of the black hole. This shift will affect the mutual information on the two sides of the black hole. The dynamics of the bulk after adding the perturbations can be conveniently  investigated in the shock wave geometry with Kruskal coordinates \cite{Dray:1984ha}.  We find that as the shift grows (more added perturbed energy at the initial time), the mutual information will be reduced more significantly, which reminds us of the phenomenon in chaos theory - butterfly effect. By turning on the graviton mass, we also find that the greater the mass is, the smaller the mutual information will be, which indicates that the spatial inhomogeneity will reduce the mutual information just like in the static case mentioned above.

This paper is organized as follows: In Section \ref{sec:shockwave} we briefly review the dynamical holographic mutual information in the shock wave geometry. We investigate the static mutual information in the massive gravity background in Section \ref{sect:holoMG}. In Section \ref{sect:dynamicMI} we study the dynamical mutual information by adding energy perturbations into the bulk of the massive gravity. Finally we draw our conclusions and discussions in Section \ref{sect:con}. Through out this paper we use natural units ($G=c=\hbar=1$) for
simplicity.

\section{Reviews of shock wave geometry and holographic mutual information }
\label{sec:shockwave}
The butterfly effect of a black hole is usually studied in the shock wave geometry, therefore, we are going to briefly introduce the key ingredients of the shock wave geometry at first. For simplicity, we will adopt a planar symmetric black hole in $3+1$ dimensions with the line element,
\begin{eqnarray}\label{peturbation}
ds^{2}=-f(r)dt^2+\frac{dr^2}{f(r)}+r^{2}(dx^2+dy^2),
\end{eqnarray}
in which $(x, y)$ are the transverse  directions in the bulk while $r$ represents the radial direction.  The Hawking temperature of this black hole is  $T=\kappa/2\pi$, in which $\kappa=f'(r)|_{r_h}/2$ is the surface gravity with $r_h$ the horizon radius. From the AdS/CFT dictionary, the temperature $T$ can be regarded as the temperature of the dual field theory.

\subsection{Shock wave geometry}
\begin{figure}
\centering
\subfigure{
\includegraphics[trim=2.3cm 3.8cm 3.7cm 3.cm, clip=true, scale=0.33]{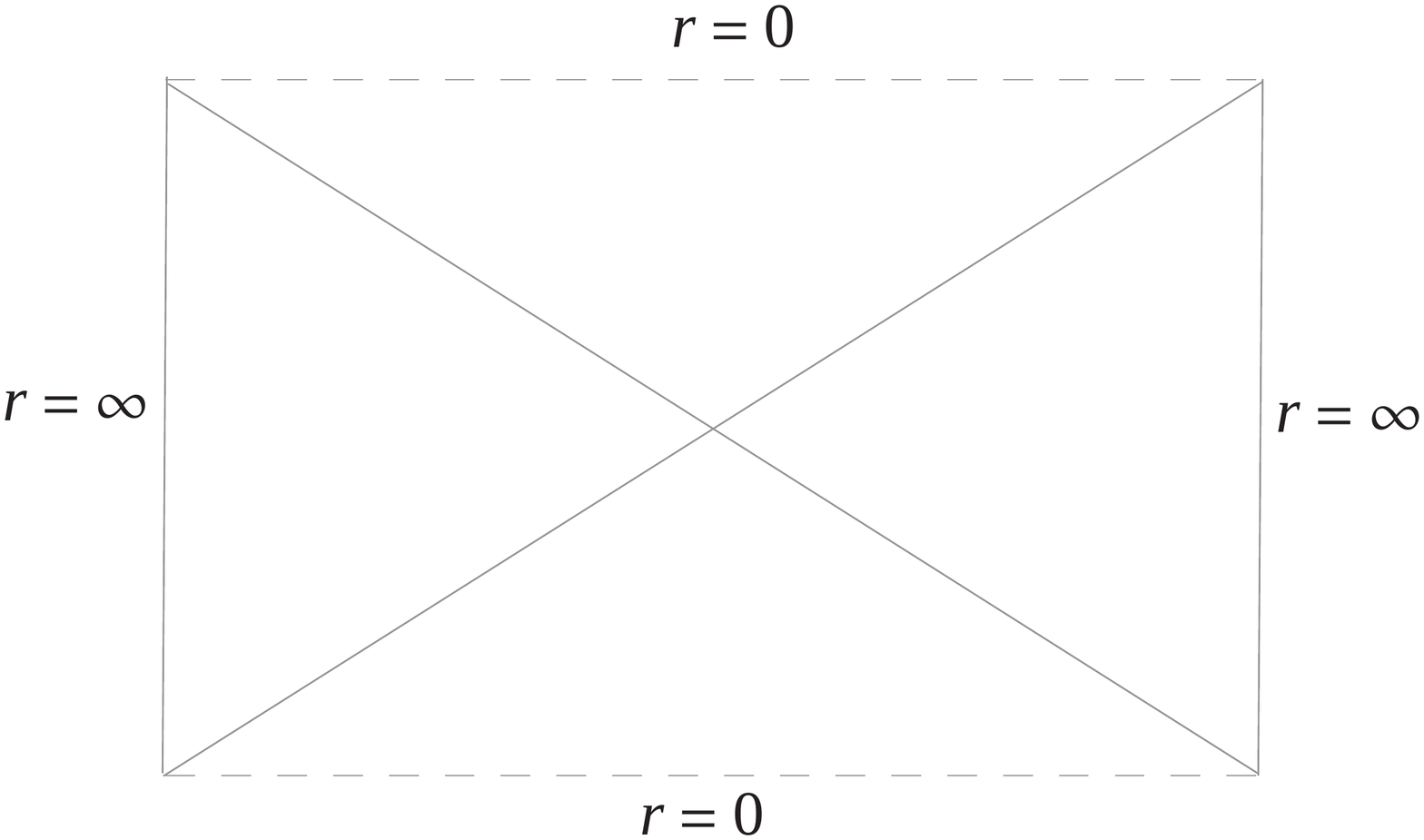}  }
\subfigure{
\includegraphics[trim=6.8cm 4.8cm 8.cm 5.3cm, clip=true, scale=0.43]{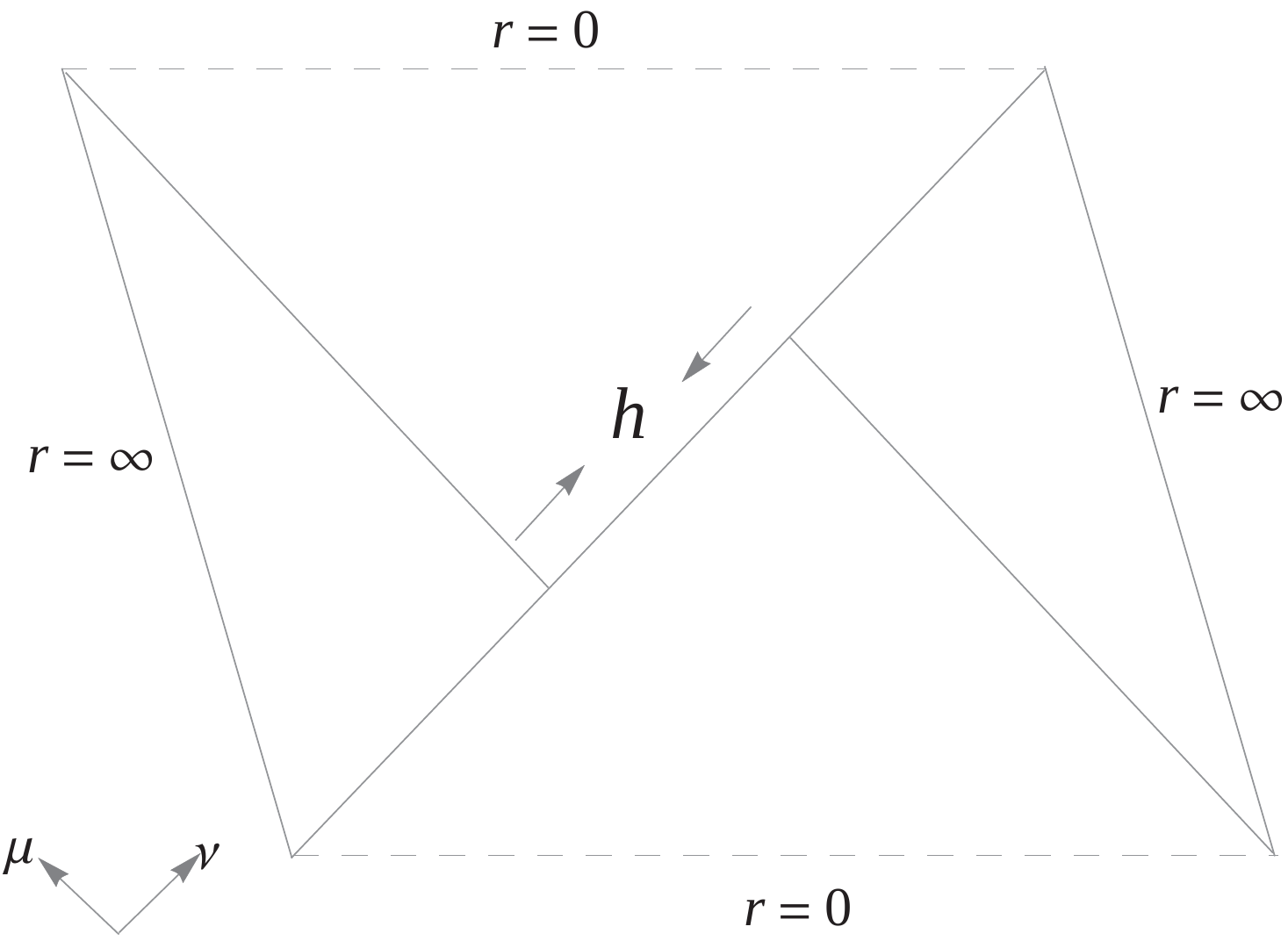}  }
 \caption{\small Penrose diagrams for an eternal black hole without (left panel) and with (right panel) a perturbation. $h$ is the shift on the horizon between the left and right Kruskal coordinate $\nu$. } \label{fig1}
\end{figure}

The shock wave geometry is conveniently  discussed in the  Kruskal coordinates \cite{Dray:1984ha}.  The metric in Eq.(\ref{peturbation}) can be rewritten as
\begin{eqnarray}\label{kruskal}
ds^{2}=\frac{1}{\kappa^2} \frac{f(r)}{\mu\nu}d \mu d\nu+r^{2}(dx^2+dy^2),
\end{eqnarray}
in which
\begin{eqnarray}
\mu=\pm e^{-\kappa U},&& \nu=\mp e^{\kappa V},\label{uv}\\
\mu\nu=-e^{2\kappa r_{\star}}, &&\mu/\nu=-e^{-2\kappa t},\label{mu}
\end{eqnarray}
where $U=t-r_{\star}$ and $V=t+r_{\star}$ are the Eddington-Finkelstein  coordinates, which can be defined by the tortoise coordinate $r_{\star}=\int dr/ f(r)$. Therefore, as $r$ approaches the event horizon and boundary, $r_{\star}$ tends to $-\infty$ and 0 respectively. The Penrose diagram of the geometry \eqref{kruskal} is shown on the left panel of Fig.\ref{fig1}, where a dot represents a two dimensional space in $(x, y)$ directions. We suppose $\mu>0, \nu<0$ in the left exterior region while $\mu<0,\nu>0$ in the right region as in \cite{Leichenauer:2014nxa}.  Thus from Eq.(\ref{mu}) we know that the event horizon and boundary are located at $\mu\nu=0$ and  $\mu\nu=-1$, respectively. The light is going along $\mu=\rm{constant}$ and $\nu=\rm{constant}$.

The shock wave geometry can be obtained by adding a small perturbation of energy $E$ into one side, for instance left side of the black hole \cite{Dray:1984ha}. Supposing that at the boundary time $t_w$ we add a light-like perturbation of the energy, which goes along a constant $\mu$ trajectory, into the left boundary. We label the Kruskal coordinates on the left side and right side as $\mu_L, \nu_L$ and  $\mu_R, \nu_R$, respectively. The constant $\mu$ trajectory of the perturbation implies
\begin{eqnarray}
\mu_L=\mu_R=e^{-\kappa t_w}.
\end{eqnarray}
In order to find the relation between $\nu_L$ and $\nu_R$, we employ the relation
\begin{eqnarray}
\mu_L\nu_L=-e^{2\kappa_L r_{ \star L}}, ~~~\mu_R\nu_R=-e^{2\kappa_R r_{ \star R}}.
\end{eqnarray}
Generally speaking, $\kappa_L=\kappa_R=\kappa$ for the energy $E$ of the perturbation is  much smaller than the mass of the black hole $M$.  On the other hand, we are interested in the case $t_w\rightarrow \infty$, which implies $r\rightarrow r_h$. In this case, we can approximate $r_{ \star}\approx\frac{1}{2 \kappa}(\log(r-r_h)+c)$ with $c$ a constant of integration.  Hence, $e^{2 \kappa r_{ \star}}=C(r-r_h)$, where $C=e^c$. Therefore, we have the identification
\be \label{vl}
\nu_L = \nu_R +C e^{\kappa t_w} (r_{hL}-r_{hR})  \equiv \nu_R + h,
\ee
 in which the relation $C_L=C_R=C$ has been used.  It should be stressed that even $(r_{hL}-r_{hR})\rightarrow 0$, the formula $e^{\kappa t_w} (r_{hL}-r_{hR})$ is still finite in \Eq{vl}. The difference $h$ between $\nu_L$ and $\nu_R$ is the shift close to the horizon. The Penrose diagram of the shock wave geometry is shown on the right panel of Fig.\ref{fig1}.

To get the  standard shock wave geometry as showed in \cite{Shenker:2013pqa}, one often employs the replacement   $\nu\rightarrow \nu+h(\theta) \Theta(\mu)$, where $\Theta(\mu)$ is  a step function. The shock wave geometry is strictly a solution to the Einstein equation $G_{\mu\nu}=\delta T_{\mu\mu}$, in which $\delta T_{\mu\mu}\sim  E e^{\frac{2\pi}{\beta} t_w}\delta (\mu) \delta (x)$ is the boost energy arising from the null perturbation at the initial time.

The scrambling time $ t_{\star}$ is defined as the value of $t_w$ when $h\sim O(1)$ for in this case the mutual information vanishes~\cite{Shenker:2013pqa}.
On the basis of  \Eq{vl} and the first law of the black hole thermodynamics, the scrambling time can be written as $
t_{\star}\sim \frac{\beta}{2 \pi}\log[ c(r_h)S$], in which $c(r_h)$ is a function of the black hole horizon. In various gravity models, this form is universal and the only difference is embodied in the function $c(r_h) $\cite{Huang2017}.



\subsection{Holographic mutual information}
As  depicted on the left panel of Fig.\ref{fig1}, an eternal black hole has  two asymptotic AdS regions, which  can be holographically described by the so-called TFD states of the two identical, non-interacting conformal field theories \cite{Maldacena:2013xja}.  Our objective is to compute the mutual information of a subregion $A$ on the left asymptotic boundary and its partner $B$ on the right asymptotic boundary. For simplicity, we will let $A=B$  so that the left and right boundaries are identical.

We are interested in  the 3+1-dimensional  planar black holes, thus the AdS boundary has a 2-dimensional space parameterized by coordinates $(x, y)$. We will consider the subregion $A$ or $B$ as a strip, which has the width $x\in(0, x_0)$ and extends along the $y$ direction with length $Y$ \footnote{Without loss of generality, we set $Y\equiv1$ in the numerics.}. Therefore, the entanglement entropy $S_A$ of the subregion $A$ is $S_A={\rm Area}_A/4$, where ${\rm Area}_A$ is the area of the minimal surface in the bulk, i.e.,
 \be \label{false}
{\rm Area}_A=\int dydx\sqrt{\gamma}=Y\int_0^{x_0} dx~ r \sqrt{f^{-1}r'^2+r^2},
\ee
where $r'=dr/dx$. If regarding the integrand in \Eq{false} as a `Lagrangian' $\mathcal{L}$, one can define
a conserved quantity associated to translation in $x$-direction, that is
\be\label{conserve}
\frac{r^3}{\sqrt{r^2+f^{-1}r'^2}}=r^2_{\rm min},
\ee
where $r_{\rm min}$ is the turning point of the surface with $r' = 0$. According to the symmetry of the surface, the turning point locates at $x=x_0/2$. With the help of the conserved equation \eqref{conserve}, $x_0$ can be written as
\be \label{theta0}
x_0 = \int_0^{x_0} dx= 2\int_{r_{\rm min}}^{\infty} \frac{dr}{r\sqrt{f}}\, \frac{1}{\sqrt{\left(r/r_{\rm min}\right)^{4}- 1}},
\ee
and the minimal area in Eq.\eqref{false} can be rewritten as
\be\label{false2}
{\rm Area}_A = 2 Y\int_{r_{\rm min}}^\infty dr~\frac{r}{\sqrt{f}}\frac{1}{\sqrt{1-(r_{\rm min}/r)^{4}}}.
\ee
Since $B$ is  identical with $A$, ${\rm Area}_B$ thus takes the same form as ${\rm Area}_A$. As stressed in the introduction, we will employ the mutual information, defined by $I(x_0)=S_A+S_B-S_{A\cup B}$, to study the correlation between regions $A$ and $B$. Therefore, our next step is to find $S_{A\cup B}$ or the area ${\rm Area}_{A\cup B}$, which is the minimal surface connecting regions $A$ (left) and $B$ (right) by passing through the horizon of the black hole. From \cite{Leichenauer:2014nxa}, the total area ${\rm Area}_{A\cup B}$ including both sides of the horizons can be expressed as
\be\label{AandB}
{\rm Area}_{A\cup B} =4Y \int_{r_{h}}^{\infty} dr~r\sqrt{f^{-1}}.
\ee
Combining all the Eqs.\eqref{mutual}, \eqref{false2} and \eqref{AandB} together, we have
\be \label{totallength}
I(x_0)=\frac14\left(4 Y\int_{r_{\rm min}}^\infty dr~\frac{r}{\sqrt{f}}\frac{1}{\sqrt{1-(r_{\rm min}/r)^{4}}}-4Y \int_{r_{h}}^{\infty} dr~\frac{r}{\sqrt{f}}\right).
\ee
We are interested in how the width of the strip $x_0$ affects the mutual information. Substituting \Eq{theta0} into \Eq{totallength}, we obtain
\be \label{totallength1}
I(x_0)=\frac12 Yx_0 r^2_{\rm min}+ Y\int_{r_{\rm min}}^\infty dr~\frac{r}{\sqrt{f}}\sqrt{1-(r_{\rm min}/r)^{4}}- Y\int_{r_{h}}^{\infty} dr~\frac{r}{\sqrt{f}}.
\ee
It is of great interest to find the critical value of the width $x_{0c}$ where the mutual information vanishes, i.e., $I(x_{0c})=0$, which leads to
\begin{equation}
x_{0c}=\frac{2}{r^2_{\rm min}}\left[ \int_{r_{h}}^{\infty} dr~\frac{r}{\sqrt{f}}- \int_{r_{\rm min}}^\infty dr~\frac{r}{\sqrt{f}}\sqrt{1-(r_{\rm min}/r)^{4}}\right].
\end{equation}

\section {Holographic mutual information in  the static background case}
\label{sect:holoMG}
Massive gravity is a deformation of the Einstein gravity with graviton mass \cite{deRham:2010ik,deRham:2010kj}. Diffeomorphism invariance is broken in massive gravity because of the graviton mass. Therefore, the stress energy tensor is not conserved any more in the dual field theory. The non-conservation of the stress energy tensor corresponds to a momentum dissipation on the dual boundary field theory \cite{Blake:2013owa}. Therefore, from this sense the graviton mass plays the role of inhomogeneity on the boundary field theory. It is of great interest to investigate the effect of inhomogeneity on the holographic mutual information, in particular, we are going to explore the effects of the graviton mass on the mutual information with and without energy perturbations in the following context. The action of the massive gravity is as follows~\cite{Vegh:2013sk,Cai:2014znn}
\begin{equation}
\label{actionmassive}
S =\frac{1}{16\pi G}\int d^{4}x \sqrt{-g} \left[ R +\frac{6}{l^2}-\frac{1}{4}F_{\mu\nu}F^{\mu\nu} +m^2 \sum^4_i c_i {\cal U}_i (g_{\mu\nu},f_{\rho\sigma})\right],
\end{equation}
where $m$ is the graviton mass parameter,  $F_{\mu\nu}=\partial_{\mu} A_{\nu}-\partial_{\nu} A_{\mu}$ is the field strength, $f_{\mu\nu}$ is the reference metric,
$c_i$ are constants,  and ${\cal U}_i$ are symmetric polynomials of the eigenvalues of the $4\times 4$ matrix ${\cal K}^{\mu}_{\ \nu} \equiv \sqrt {g^{\mu\alpha}f_{\alpha\nu}}$:
\begin{eqnarray}
\label{eq2}
&& {\cal U}_1= [{\cal K}], \nonumber \\
&& {\cal U}_2=  [{\cal K}]^2 -[{\cal K}^2], \nonumber \\
&& {\cal U}_3= [{\cal K}]^3 - 3[{\cal K}][{\cal K}^2]+ 2[{\cal K}^3], \nonumber \\
&& {\cal U}_4= [{\cal K}]^4- 6[{\cal K}^2][{\cal K}]^2 + 8[{\cal K}^3][{\cal K}]+3[{\cal K}^2]^2 -6[{\cal K}^4].
\end{eqnarray}
The square root in ${\cal K}$ stands for $(\sqrt{A})^{\mu}_{\ \nu}(\sqrt{A})^{\nu}_{\ \lambda}=A^{\mu}_{\ \lambda}$ and $[{\cal K}]=K^{\mu}_{\ \mu}=\sqrt {g^{\mu\alpha}f_{\alpha\mu}}$. In 3+1-dimensions, a black hole solution with line element in Eq.\eqref{peturbation} can have the gravitational potential as
\begin{equation}
f(r)=-\frac{ M}{r}+\frac{Q^2}{4 r^2}+\frac{r^2}{l^2}+\frac{c_0 c_1 m^2}{2} r+c_0^2 c_2 m^2,
 \end{equation}
in which  $M$ and $Q$ are the mass and charge of the black hole respectively; $c_0, c_1$ and $c_2$ are constant parameters associated to the graviton mass. In this paper, we will fix the parameters $c_0=c_1=1, c_2=-1/2$ in order to render the background thermodynamically stable \cite{Cai:2014znn,Hu:2015dnl}. The temperature of the black hole  is readily obtained as
\be\label{temMG}
T_{MG}=\frac{3}{4 \pi }-\frac{Q^2}{16 \pi }+\frac{c_0^2 c_2 m^2}{4 \pi
   }+\frac{c_0 c_1 m^2}{4 \pi
   }.
\ee
\subsection{Effect of the width of the strip on the  mutual information}
\label{subsect:width}
\begin{figure}[h]
\centering
\subfigure{
\includegraphics[trim=8.5cm 6.5cm 8.5cm 6.5cm, clip=true, scale=0.55]{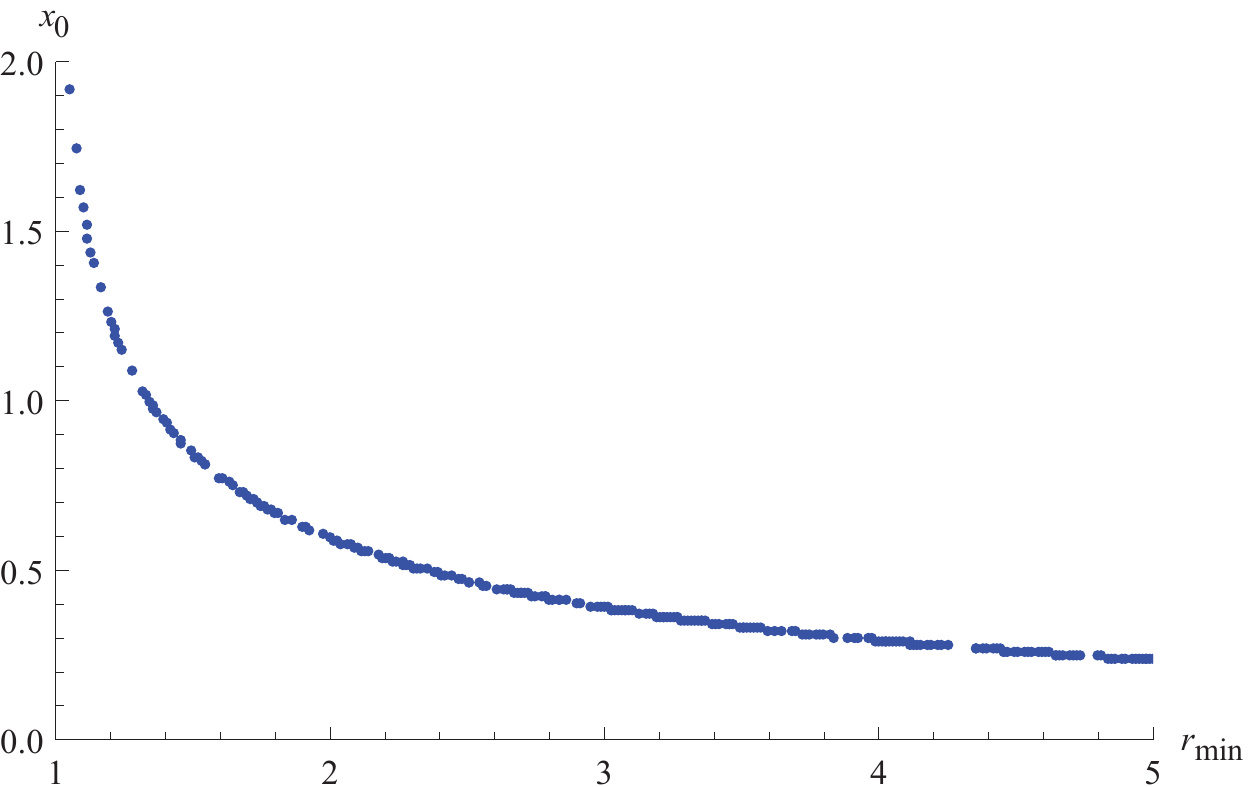}  }
\subfigure{
\includegraphics[trim=8.5cm 6.5cm 8.5cm 6.5cm, clip=true, scale=0.55]{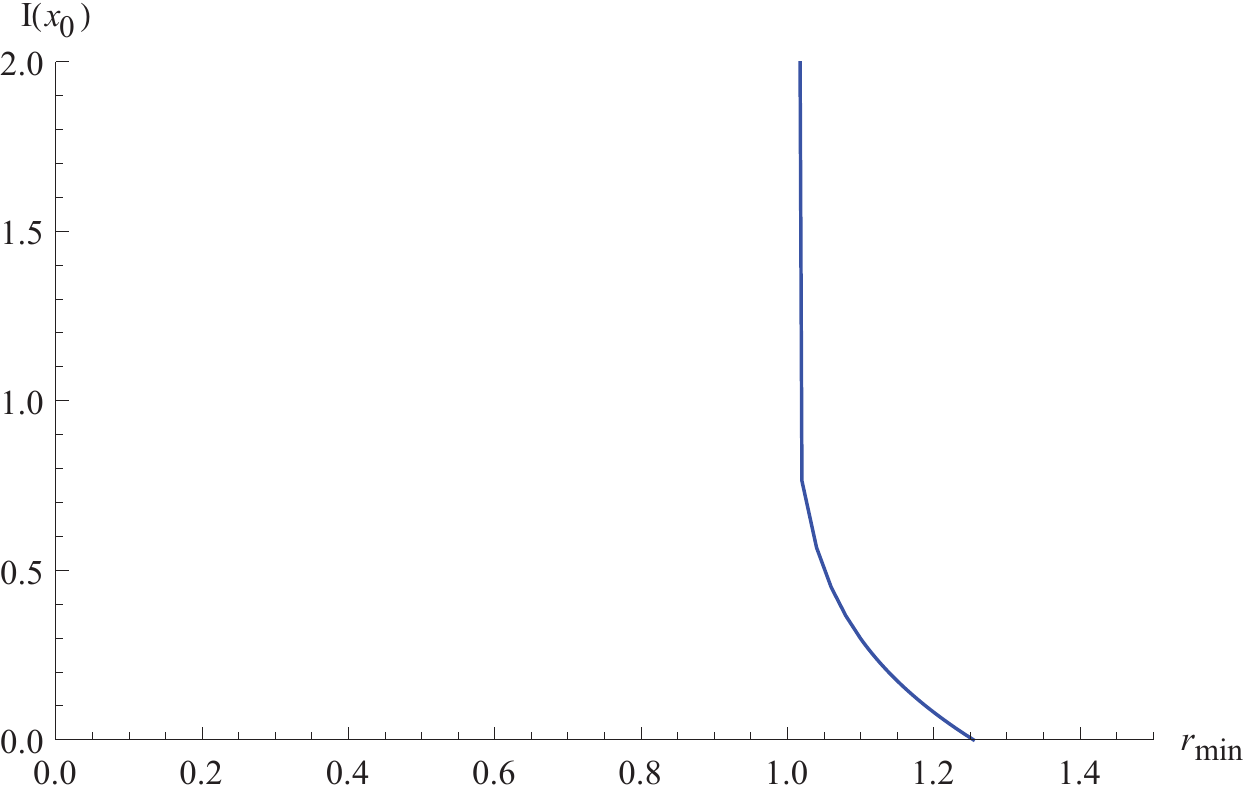}  }
 \caption{\small Left:  The relation between the width of the strip $x_{0}$ and $r_{\rm min}$.  Right: The relation between the mutual information $I(x_{0})$ and the position of the turning point $r_{\rm min}$. For both cases, we set $m= 0.6$, $Q=2$, and $r_h=1$. } \label{addfig2}
\end{figure}
\begin{figure}[h]
\centering
\subfigure{
\includegraphics[trim=8.5cm 6.4cm 8.5cm 6.4cm, clip=true, scale=0.55]{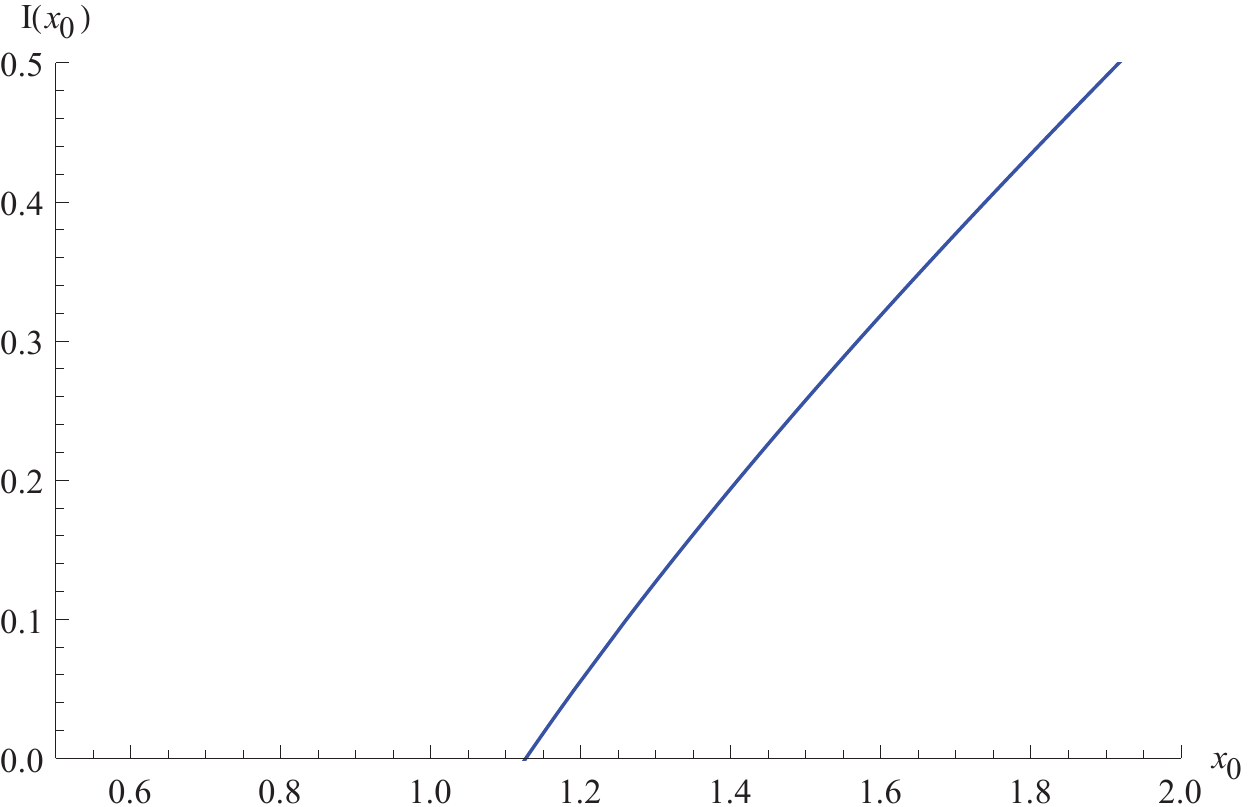}  }
 \caption{\small The relation between the mutual information $I(x_{0})$ and width of the strip $x_0$ for  $m= 0.6$, $Q=2$, and $r_h=1$. } \label{addfig3}
\end{figure}

Firstly, we are going to study the relation between the mutual information and the width of the strip in the background of massive gravity. One can readily read off the relation between the width of the strip and the position of the turning point $r_{\rm min}$ from  Eq.\eqref{theta0}. The relation is shown on the left panel of Fig.\ref{addfig2}. From Eq.\eqref{theta0} one finds that as $r_{\rm min}\to r_h$, the integral for the width $x_0$ diverges, which can be seen as well from the left plot of Fig.\ref{addfig2}. Intuitively the left panel of Fig.\ref{addfig2} is also correct, since as we know greater $r_{\rm min}$ is closer to the infinite boundary, therefore, it is obvious that the greater $r_{\rm min}$ corresponds to the smaller width of the strip. From  Eq.\eqref{totallength} or \eqref{totallength1}, one can see that as $r_{\rm min}\to r_h$ the mutual information diverges, since in this case the widths of the strips on the two boundaries are nearly divergent (cf. the left panel of Fig.\ref{addfig2}). From \cite{Leichenauer:2014nxa} we know that the mutual information for divergent strips will be divergent too. This phenomenon can also be found on the right panel of  Fig.\ref{addfig2}. In addition, from this subgraph we observe that the mutual information vanishes where $r_{\rm min}\approx1.25$, which indicates that there is a critical value for the position of the turning point (or equivalently a critical width of the strip $x_{0c}$ from the left panel of Fig.\ref{addfig2}) rendering the mutual information vanishing. Combining the two panels in Fig.\ref{addfig2}, we plot the relation between the mutual information and the width of the strip $x_0$ explicitly in Fig.\ref{addfig3}, from which we can clearly see that the critical width of the strip is roughly $x_{0c}\approx1.13$. In addition, we find that the mutual information always grows as the width of the strip increases, which is easy to understand  for in this case the subsystems on the two asymptotic boundaries are larger and their entanglements becomes greater as well.

\subsection{Effects of the graviton mass and black hole charge on the mutual information}
\label{subsect:IMQ}
\begin{figure}[h]
\centering
\subfigure{
\includegraphics[trim=8.5cm 6.4cm 8.5cm 6.4cm, clip=true, scale=0.55]{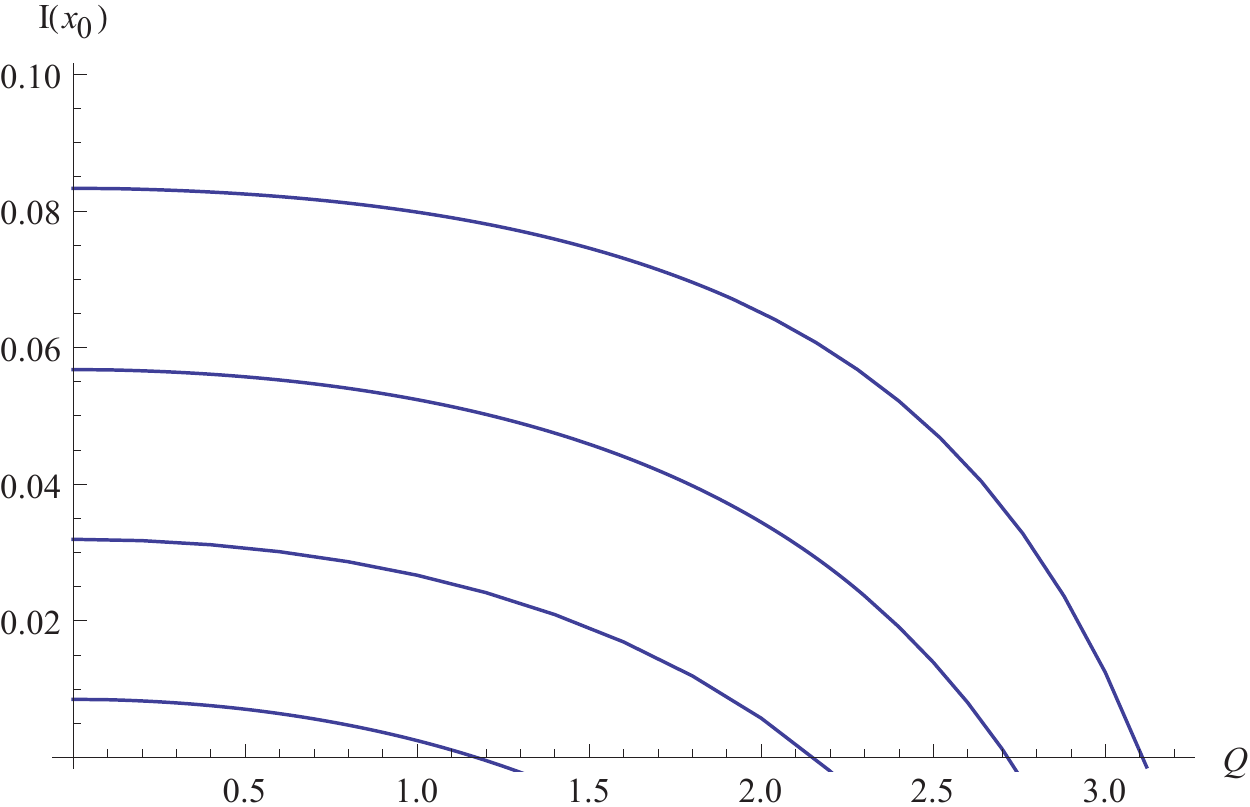}  }
\subfigure{
\includegraphics[trim=8.5cm 6.4cm 8.5cm 6.4cm, clip=true, scale=0.55]{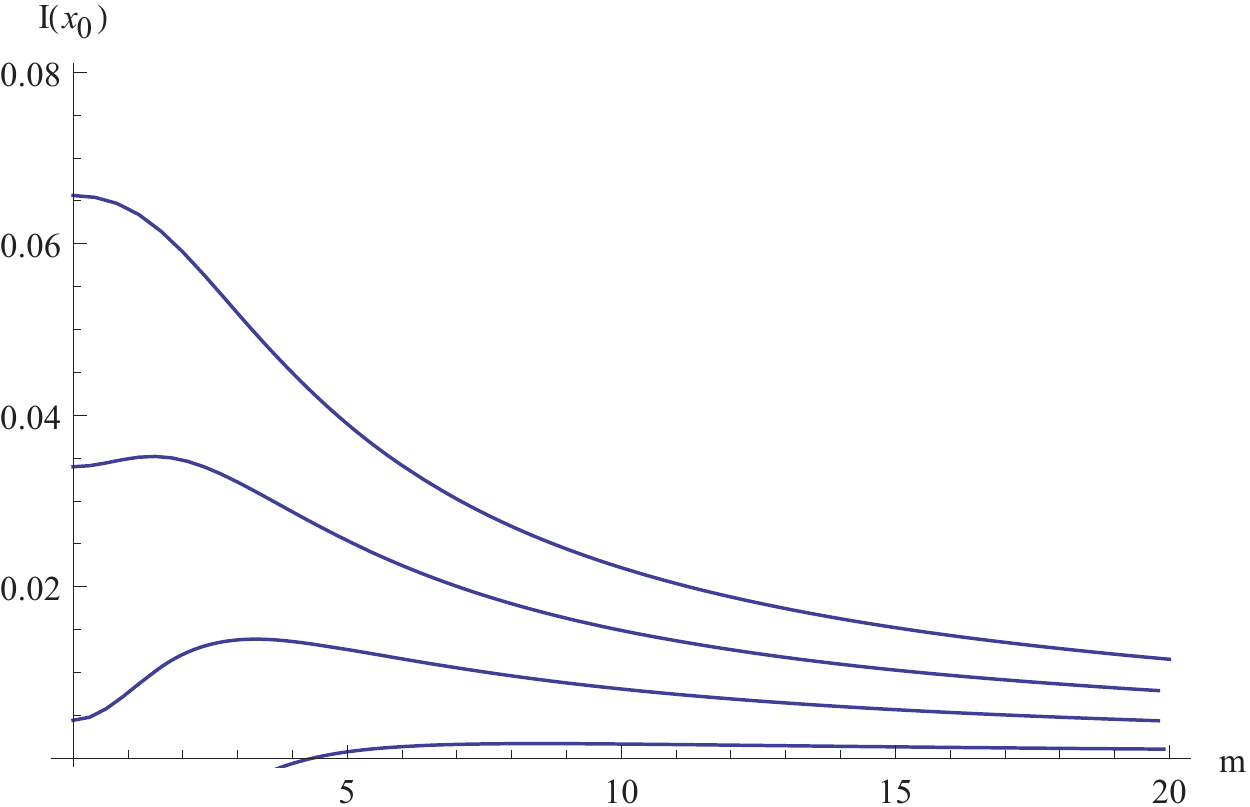}  }
 \caption{\small  Left:  The relation between  $I(x_0)$ and black hole charge $Q$ while fixing $m=0.6$, $r_h$=1.  The curves from top to down correspond to $r_{\rm min}$ increasing from {1.21 to 1.27} with steps 0.02; Right: The relation between  the mutual information $I(x_0)$ and the graviton mass $m$ by fixing $Q=2$,  $r_h$=1. Curves from top to down correspond to $r_{\rm min}$ increasing from { 1.21 to 1.27 with steps 0.02}. } \label{fig7}
\end{figure}
The effects of graviton  mass $m$ and charge $Q$ of the black hole on the mutual information $I(x_0)$ are shown in Fig.\ref{fig7}. From the left panel of Fig.\ref{fig7} we see that for each curve, the mutual information decreases as the charge increases. Besides, there is a critical charge $Q_c$ that makes the mutual information vanishing, which means that there is no entanglement between the paired subregions we considered. For a small fixed charge, we find that the mutual information is smaller for greater $r_{\rm min}$. As we know from the preceding subsection, bigger $r_{\rm min}$ actually corresponds to smaller width of the strip on the boundary. Therefore, the left panel of Fig.\ref{fig7} also indicates that smaller subregions have smaller mutual information between them, which is consistent with the preceding subsection.

From Ref.~\cite{Leichenauer:2014nxa} we learn that when the temperature of the black hole grows, the critical width of the strip decreases. It also means that for a fixed width of the strip, as temperature grows the mutual information of the two strips will grow as well. In the case of the massive gravity, the temperature will decrease as the charge $Q$ increases when fixing other parameters of the background, please refer to Eq.\eqref{temMG}. Therefore, from the left panel of Fig.\ref{fig7} we see that as charge grows (temperature decreases) the mutual information decreases monotonically, which matches the conclusions mentioned before. Incidentally, we also checked other cases of the graviton mass, similar results were obtained.

However, it is interesting to see that the mutual information does not have monotonic decreasing behavior to the graviton mass on the right panel of Fig.\ref{fig7}. In particular, when $r_{\rm min}$ is bigger ($x_0$ is smaller) the mutual information first increases to a maximum value and then decreases with respect to the graviton mass. Let's call the critical graviton mass as $m_c$ which corresponds to the maximum value of mutual information. From  Eq.\eqref{temMG} we see that as $m$ grows the temperature of the black hole grows as well when fixing other parameters. Therefore, from the conclusions in the preceding subsection it seems that as $m$ grows the mutual information should increase as well, since the mutual information increases with respect to the temperature. However, as we see from the right panel of Fig.\ref{fig7}, the mutual information decreases as $m$ grows in the regime $m>m_c$, which contradicts the above conclusions.

The possible way coming to rescue is that the conclusion in the previous subsection is mainly  valid in the (near-)homogeneous case. However, when taking into account of the inhomogeneous effects induced by the graviton mass, the conclusions should be different. Therefore, we can infer from the right panel of Fig.\ref{fig7} that when $m<m_c$ the inhomogeneous effects are tiny. Hence, in this case the mutual information still grows as the temperature grows (i.e., $m$ increases); However, when $m>m_c$ the inhomogeneous effects are significant, which will disrupt the mutual information and finally render it vanishing as $m$ is big enough. Therefore, we can infer that greater inhomogeneity will spoil the mutual information.

Finally, let's come to the top curve on the right panel which shows that the mutual information decreases monotonically with respect to $m$. In fact the top curve corresponds to a smaller $r_{\rm min}$ (bigger width $x_0$ of the strip). Therefore, it makes us  speculate that the inhomogeneity will have greater effects on longer strips than shorter ones to reduce the mutual information. Therefore, for a long strip the mutual information only decreases according to $m$ since now the inhomogeneous effects dominate.

\subsection{The critical
 width }

\begin{figure}[ht]
\centering
\subfigure{
\includegraphics[trim=8.5cm 6.35cm 8.5cm 6.35cm, clip=true, scale=0.55]{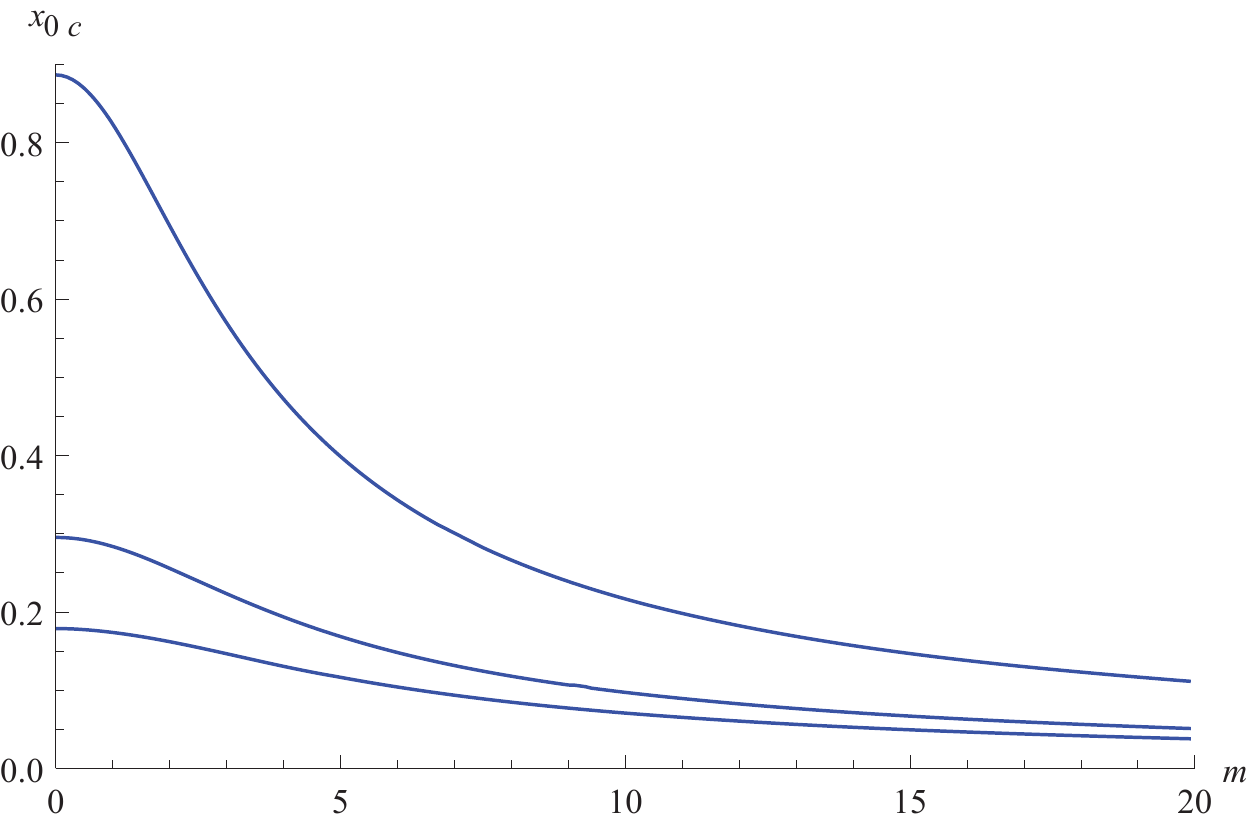}  }
\subfigure{
\includegraphics[trim=8.5cm 6.35cm 8.5cm 6.35cm, clip=true, scale=0.55]{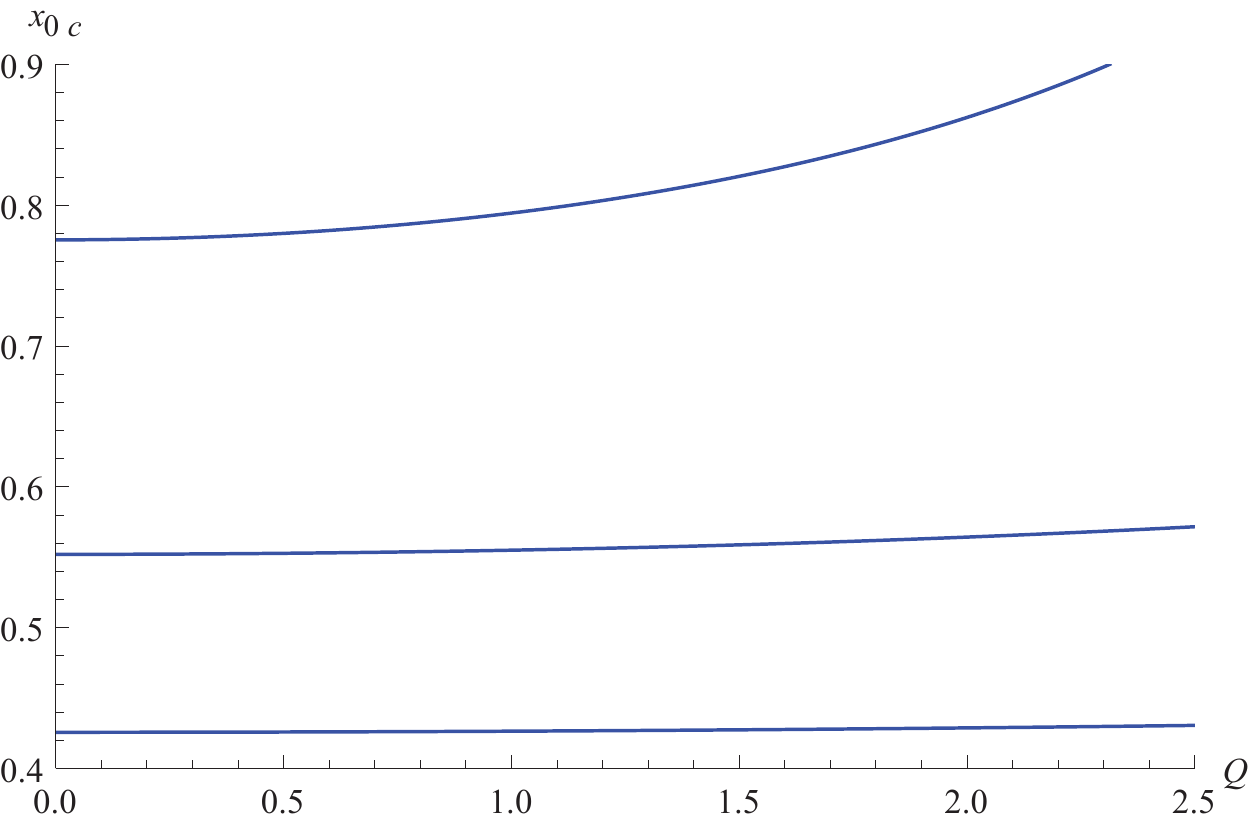}  }
 \caption{\small Left: Relation between the critical width  $x_{0c}$ and graviton mass $m$ by fixing charge $Q=2$. Curves from top to down correspond to $r_h$ increasing from 1 to 5 with steps 2, respectively. Right: Relation between critical width $x_{0c}$ and the charge $Q$ while fixing the graviton mass $m=0.6$. Curves from top  to down correspond to $r_h$ increasing from 1 to 2 with steps 0.4, respectively.} \label{fig8}
\end{figure}
The critical width $x_{0c}$ of the strip is the width that renders the mutual information vanishing, i.e., $I(x_{0c})=0$. On the left panel of Fig.\ref{fig8}, by fixing the charge $Q=2$, we see that the critical width $x_{0c}$ decreases as the graviton mass $m$ grows for various black hole horizons $r_h$. This means that the greater inhomogeneity will reduce the critical width $x_{0c}$. For a fixed value of $m$, the critical width $x_{0c}$ decreases with respect to the increasing horizon $r_h$. We know that the greater horizon corresponds to higher temperature of the black hole, therefore, the left plot of Fig.\ref{fig8} also indicates that the higher temperature also reduces the critical width $x_{0c}$.

On the right panel of Fig.\ref{fig8}, we plot the relation between the critical width $x_{0c}$ and the charge $Q$ for various horizons $r_h$. For a fixed horizon, we find that when $Q$ increases the critical width $x_{0c}$ increases as well. In particular, when $r_h$ is small the critical width increases more rapidly; However, when $r_h$ is big, $x_{0c}$ increases more mildly. As we know from Eq.\eqref{temMG} that as $Q$ increases the temperature $T_{MG}$ decreases. Therefore, the right plot of Fig.\ref{fig8} indicates that as temperature of the black hole decreases the critical width $x_{0c}$ will increase, which is consistent with the previous analysis. Moreover, for a fixed charge $Q$, the critical width $x_{0c}$ decreases as $r_h$ increases, which states that the higher temperature reduces the critical width $x_{0c}$. This statement matches the conclusions above.

\section{ Holographic mutual information in the dynamical background case}
\label{sect:dynamicMI}
As we know from Section \ref{sec:shockwave}, when a small perturbation is added from the left boundary to the bulk,  there will be a shift $h(x)$ near the horizon in the $\nu$ direction for a long enough time $t_w$. A shock wave geometry thus forms and the wormhole connecting the left and right regions may be destroyed in some circumstances. Therefore, the mutual information between the two subregions will be disrupted. In the previous section we studied the holographic mutual information in the static background of massive gravity; In this section, we are going to investigate the dynamical behavior of the holographic mutual information in the shock wave background of massive gravity.

\begin{figure}
\centering
\subfigure{
\includegraphics[trim=5.8cm 21cm 7cm 4.5cm, clip=true, scale=1]{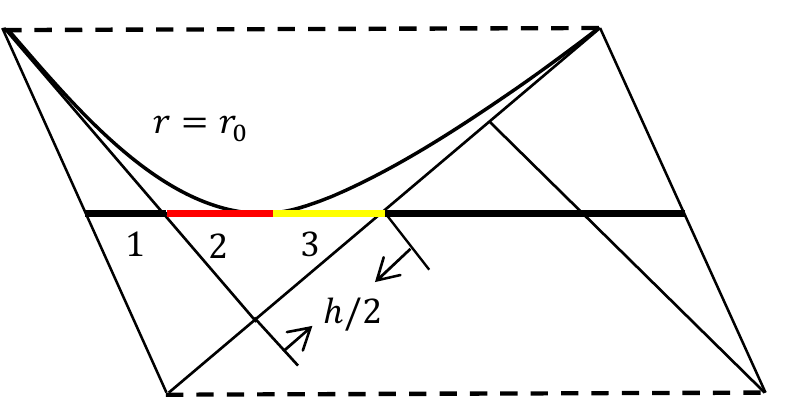}  }
 \caption{\small The Penrose diagram of the shock wave geometry. The horizontal colorful line is the minimal surface connecting one of the two ends of the strips (one dot in the line represents a two dimensional surface spanned by $(x, y)$ coordinates). The left half of the surface is divided  into three segments, i.e., black line, red line and yellow line. The surface $r=r_0$ separates the line 2 and 3. } \label{fig9}
\end{figure}

As in Section \ref{sect:holoMG}, we suppose that the strip $A$ sits in the left asymptotic boundary and its identical partner $B$ in the right boundary. After adding the light-like perturbations, the areas of minimal surfaces ${\rm Area}_A$ and ${\rm Area}_B$ are unaffected by the shock wave since they do not cross the horizon, while the ${\rm Area}_{A\cup B}$ is affected by the shock wave since it passes across the horizon and stretches through the wormhole. The sketchy plot of the surface ${\rm Area}_{A\cup B}$ is shown in Fig.\ref{fig9}. Our main goal thus is to calculate the ${\rm Area}_{A\cup B}$ and to study how it changes with respect to the shift $h(x)$ for a fixed boundary separation.

Because of the symmetry of the minimal surface, we should only calculate the areas for the regions 1, 2 and 3 in Fig.\ref{fig9} \cite{Leichenauer:2014nxa,Sircar:2016old}.  At a surface with constant $x$, the induced metric  can be written as
 \begin{eqnarray}
ds^{2}=[-f(r)+\frac{1}{f(r)}\dot{r}^2]dt^2+r^{2}dy^2,
\end{eqnarray}
 in which $\dot{r}=dr/dt$. The area of minimal surface for the regions 1, 2 and 3 in Fig.\ref{fig9}  is then given by
\be \label{ab}
\overline{\rm Area}_{A\cup B}(h) = \int dt~r\sqrt{-f  +f^{-1}\dot{r}^2}.
\ee
If regarding the integrand  in \Eq{ab} as a `Lagrangian' $\mathcal{L}$, we can define a conserved `Hamiltonian' $\mathcal{H}$ as
\be \label{aab}
\mathcal{H}= \frac{-rf }{\sqrt{-f + f^{-1}\dot{r}^2}} =  r_0\sqrt{-f_0},
\ee
in which $f_0= f(r_0)$ and  $r_0$  is  the radial position behind the horizon satisfying $\dot{r}=0$. From  \Eq{aab}, we know that as  $r_0\to r_h$,  $\mathcal{H}$ tends to $0$,  which corresponds to the case that the shock wave is absent for $h \rightarrow 0$.
From the conservation equation Eq.\eqref{aab}, $t$ coordinate can be written as a function of $r$
\be\label{eqtime}
t(r) =\pm \int \frac{dr}{f\sqrt{1 +\mathcal{H}^{-2}fr^2}},
\ee
where $\pm$  denote  $\dot{r}>0$ and $\dot{r}<0$ respectively.  Substituting \Eq{eqtime} into \Eq{ab}, we can get a time independent integral
\be \label{}
\overline{\rm Area}_{A\cup B}(h) = \int dr~\frac{r^2}{\sqrt{\mathcal{H}^2+fr^2}}.
\ee
Therefore, the area of regions 1+2+3 in Fig.\ref{fig9} can be rewritten as
\be \label{}
\overline{\rm Area}_{A\cup B}(h) = \int_{r_h}^{\infty} dr~\frac{r^2}{\sqrt{\mathcal{H}^2+fr^2}}+2 \int_{r_0}^{r_h} dr~\frac{r^2}{\sqrt{\mathcal{H}^2+fr^2}}.
\ee
The second term contains a prefactor 2 stemming from the fact that  the second and third regions have the same area.   The total area ${\rm Area}_{A\cup B}(h)$ which connects the left and right boundaries thus is
\be \label{}
{\rm Area}_{A\cup B}(h) = 2\int_{r_h}^{\infty} dr~\frac{r^2}{\sqrt{\mathcal{H}^2+fr^2}}+4 \int_{r_0}^{r_h} dr~\frac{r^2}{\sqrt{\mathcal{H}^2+fr^2}}.
\ee
It should be stressed that the first segment contains a divergent $h$-independent term which must be subtracted by a pure AdS contribution as we compute it numerically. Considering the contribution of ${\rm Area}_A$ and  ${\rm Area}_B$, the mutual information in the shock wave geometry can be expressed as
\be \label{ih}
I(h, x_0) = \frac{Y}{4}\left(4 \int_{r_{\rm min}}^{\infty}dr~\frac{r}{\sqrt{f}}\frac{1}{\sqrt{1-(r_{\rm min}/r)^{4}}}-4\int_{r_h}^{\infty} dr~\frac{r^2}{\sqrt{\mathcal{H}^2+fr^2}}-8 \int_{r_0}^{r_h} dr~\frac{r^2}{\sqrt{\mathcal{H}^2+fr^2}}\right).
\ee
 We are going to find the relation between  $I(h, x_0)$ and $h$. We see that $I(h, x_0)$ depends on the location of $r_0$ for a fixed $r_h$.   Thus in order to proceed, we should find the relation between  $h$ and $r_0$.
 From Fig.\ref{fig9}, the first segment goes from the boundary at $(\mu,\nu) = (1,-1)$ to $(\mu,\nu) = (\mu_1,0)$, in which
\be
\mu_1 =  \exp[-\kappa \int_{r_{h}}^{\infty} \frac{dr}{f} (1- \frac{1}{\sqrt{1 +\mathcal{H}^{-2}fr^2}})].
\ee
where we have used \Eq{uv}. The second segment stretches from $(\mu_1,0)$ to  $(\mu_2,\nu_2)$ at the surface $r = r_0$. The coordinate $\mu_2$ can be determined by the relation
\be
\frac{\mu_2}{\mu_1} =  \exp[-\kappa \int_{r_{0}}^{r_h} \frac{dr}{f} (1- \frac{1}{\sqrt{1 +\mathcal{H}^{-2}fr^2}})].
\ee
 In contrast, the coordinate $\nu_2$ can be determined by choosing a reference surface $r=\bar{r}$ for  which $r_{\star}=0$ in the black hole  interior. Thus, we reach
\be
\nu_2 = \frac{1}{\mu_2} \exp(2 \kappa \int_{\bar{r}}^{r_0} \frac{dr}{f}).
\ee
The third segment stretches from $(\mu_2,\nu_2)$ to  $(\mu_3=0,\nu_3=h/2)$. Therefore, with the relation
\be
\frac{\nu_3}{\nu_2} = \frac{h}{2 \nu_2} =\exp[\kappa \int_{r_{0}}^{r_h} \frac{dr}{f} (1- \frac{1}{\sqrt{1 +\mathcal{H}^{-2}fr^2}})]=\frac{\mu_1}{\mu_2},
\ee
we can express $h$ as
\be \label{h}
h=2 \exp(\Xi_1+\Xi_2+\Xi_3),
\ee
where
\begin{eqnarray}
\Xi_1&=& 2 \kappa \int_{\bar{r}}^{r_0} \frac{dr}{f},\\
\Xi_2&=&2 \kappa \int_{r_{0}}^{r_h} \frac{dr}{f} (1- \frac{1}{\sqrt{1 +\mathcal{H}^{-2}fr^2}}),\\
\Xi_3&=& \kappa \int^{\infty}_{r_h} \frac{dr}{f} (1- \frac{1}{\sqrt{1 +\mathcal{H}^{-2}fr^2}})\label{xi3}.
\end{eqnarray}
It is obvious that $h$ depends on the location of $r_0$ for a fixed $r_h$. Combining \Eq{ih} with \Eq{h}, one can get the relation between $I(h, x_0)$ and $h$. Next, we will study the relation between $I(h, x_0)$ and $h$ numerically in the background of the shock wave geometry in the massive gravity.

\subsection{Black hole charge and butterfly effect}
\begin{figure}[h]
\centering
\subfigure[]{
\includegraphics[trim=8.5cm 6.35cm 8.5cm 6.35cm, clip=true, scale=0.55]{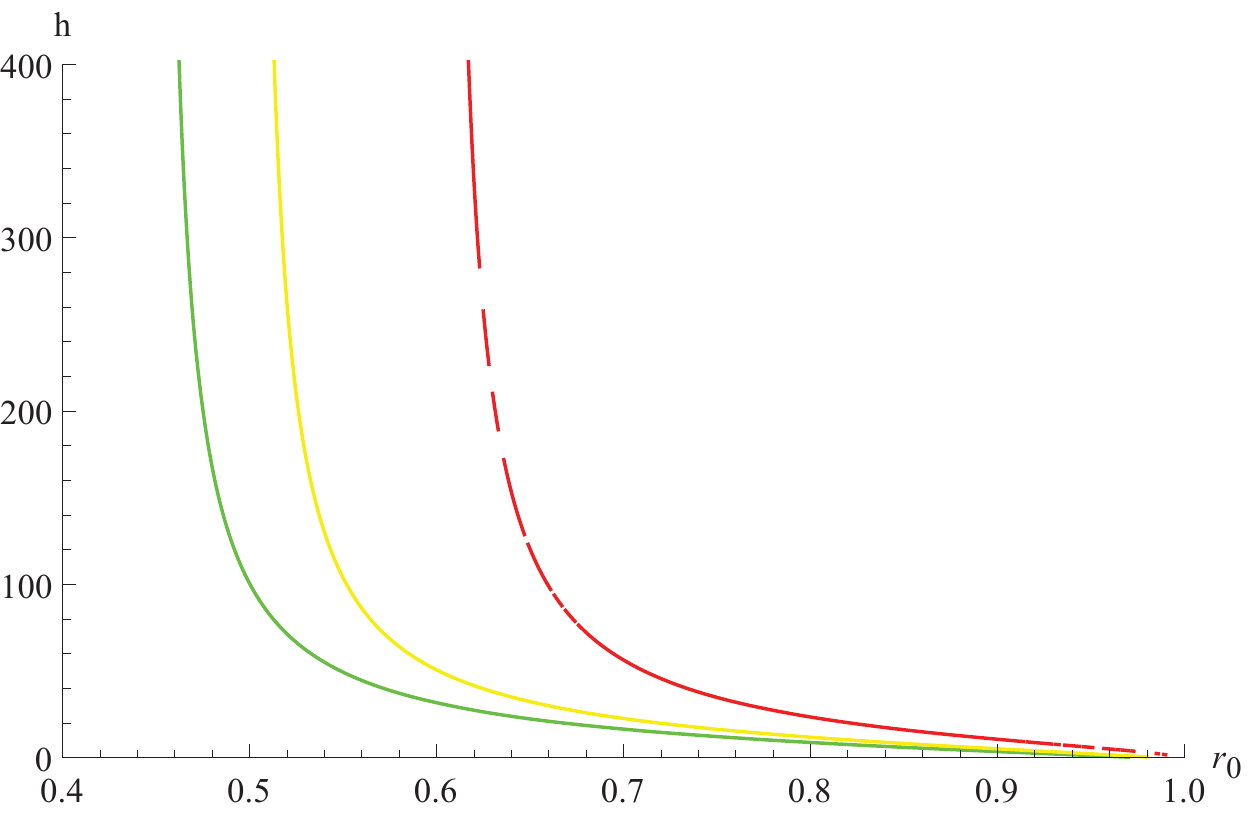}  }
\subfigure[]{
\includegraphics[trim=8.5cm 6.35cm 8.5cm 6.35cm, clip=true, scale=0.55]{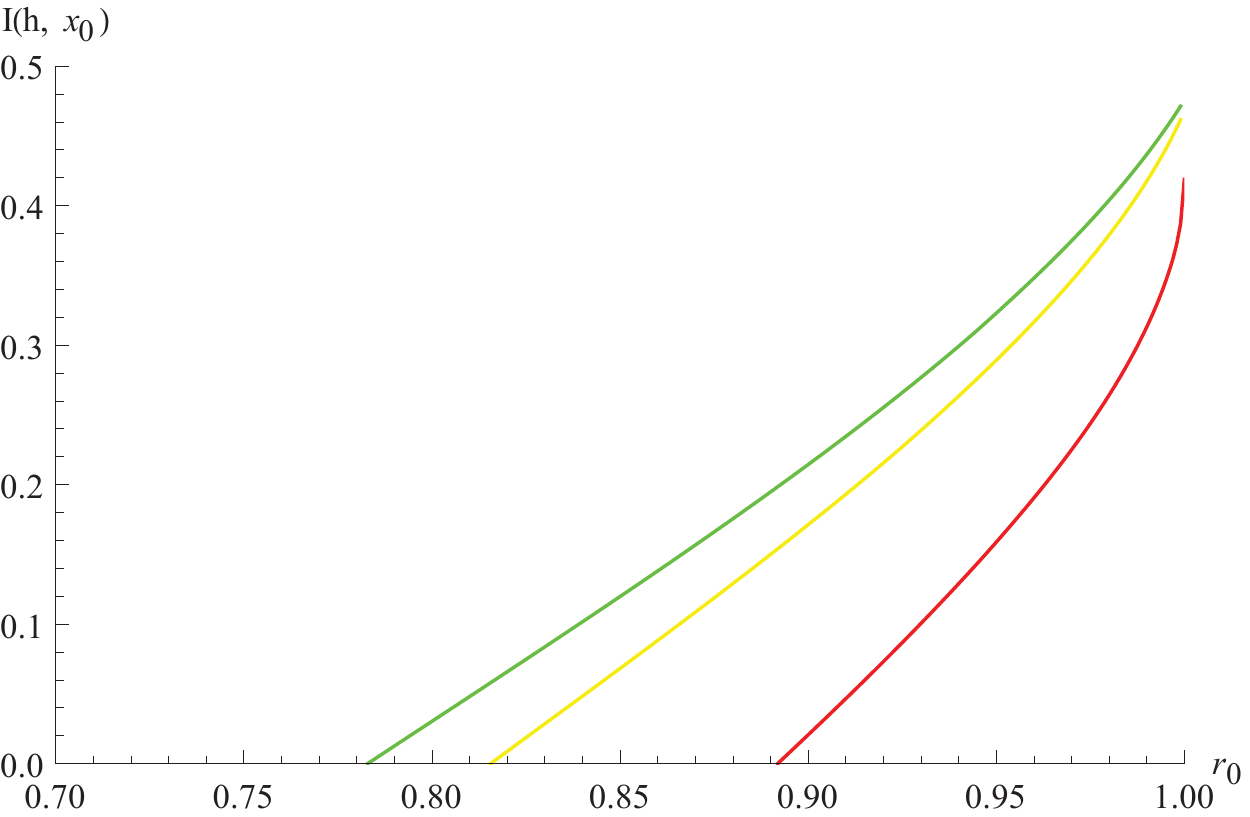}  }\\
\subfigure[]{
\includegraphics[trim=8.5cm 6.35cm 8.5cm 6.35cm, clip=true, scale=0.55]{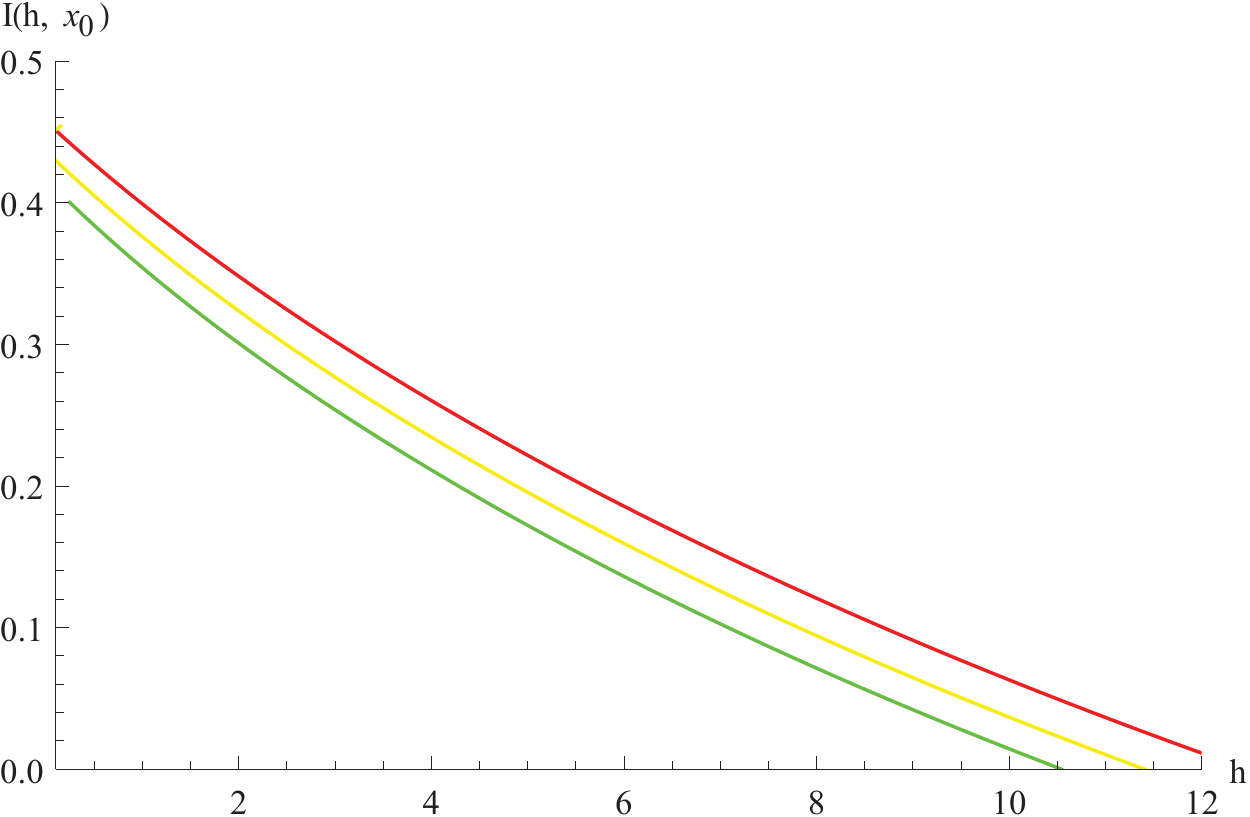}  }
 \caption{\small Panel (a): The relation between the shift $h$ and the minimal radius $r_0$ for various charges $Q$; Panel (b): The relation between the mutual information $I(h, x_0)$ and  $r_{0}$  for different $Q$; Panel (c):  The relation between $I(h, x_0)$ and  $h$  for various $Q$. In numerics, we have set  $ \bar{r}=0.2$, $r_{min} = 50$, $m=0.6, r_h=1$. The red, yellow and green lines correspond to $Q=0.6, 0.7, 0.8$ respectively.  } \label{fig16}
\end{figure}
We are going to study how the charge affects the shock wave geometry and the dynamical behavior of mutual information in the following context. As we already knew from the previous subsection that as $r_0\to r_h$ (we set $r_h=1$ in numerics), the shift $h$ vanishes, which can also be found in the plot (a) of Fig.\ref{fig16}. The vanishing of $h$ means there is no added perturbation into the bulk, therefore, the shock wave geometry goes back to the unperturbed or the static case which we studied in the previous section.

From the formula of $\Xi_3$ in Eq.\eqref{xi3}, we can see that if the denominator in the integrand vanishes, i.e.,
$
\sqrt{1 +\mathcal{H}^{-2}fr^2}=0,
$
the shift $h$ diverges. Since $\mathcal{H}$ is a conserved quantity, the above equation can be readily transformed to
\be\label{rcrit}
\frac{d(fr^2)}{dr}\bigg|_{r=r_{\rm crit}}=f'(r_{\rm crit})r_{\rm crit}+2f(r_{\rm crit})=0,
\ee
 in which, $r_{\rm crit}$ is the critical position that makes the shift $h$ divergent.  For instance, in the case of $Q=0.6$ and $m=0.6$, a physical solution to  Eq.\eqref{rcrit} is $r_{\rm crit}\approx0.6275$ which is consistent with  plot (a) of Fig.\ref{fig16}. We have also checked the correctness of  Eq.\eqref{rcrit} for other parameters.

The mutual information grows as $r_0$ increases, which is shown on  panel (b) of Fig.\ref{fig16}. From this plot we see that there are also critical values of $r_0$ that renders mutual information vanishing. In fact, one can see from  plot (a) of Fig.\ref{fig16} that for a fixed $r_0$, greater values of the charge correspond to smaller values of shift $h$. As we know that the shift $h$ is proportional to the energy of the added perturbation, thus we can deduce from  panel (b) of Fig.\ref{fig16} that the more the added energy perturbation is, the smaller the mutual information will be. The relationship between mutual information and the shift $h$ can be more clearly seen from the panel (c) of Fig.\ref{fig16}. As the shift grows to a critical value $h_{\rm crit}$, the mutual information will vanish, i.e., when the added perturbation is big enough, it will finally disrupt the entanglement between the two strips. The disruption of the mutual information depending on the added perturbation in the initial time reminds us of the phenomenon in chaos theory, i.e., {\it butterfly effect}. For a fixed value of shift, the mutual information decreases as the charge grows, which is consistent with the statements in the previous section. In particular, when the shift is zero (or equivalently $r_0\to 1$), i.e., there is no added perturbation into the bulk, it will go back to the static case discussed in the previous section.

\subsection{Graviton mass and butterfly effect}
\begin{figure}[h]
\centering
\subfigure[]{
\includegraphics[trim=8.5cm 6.35cm 8.5cm 6.35cm, clip=true, scale=0.55]{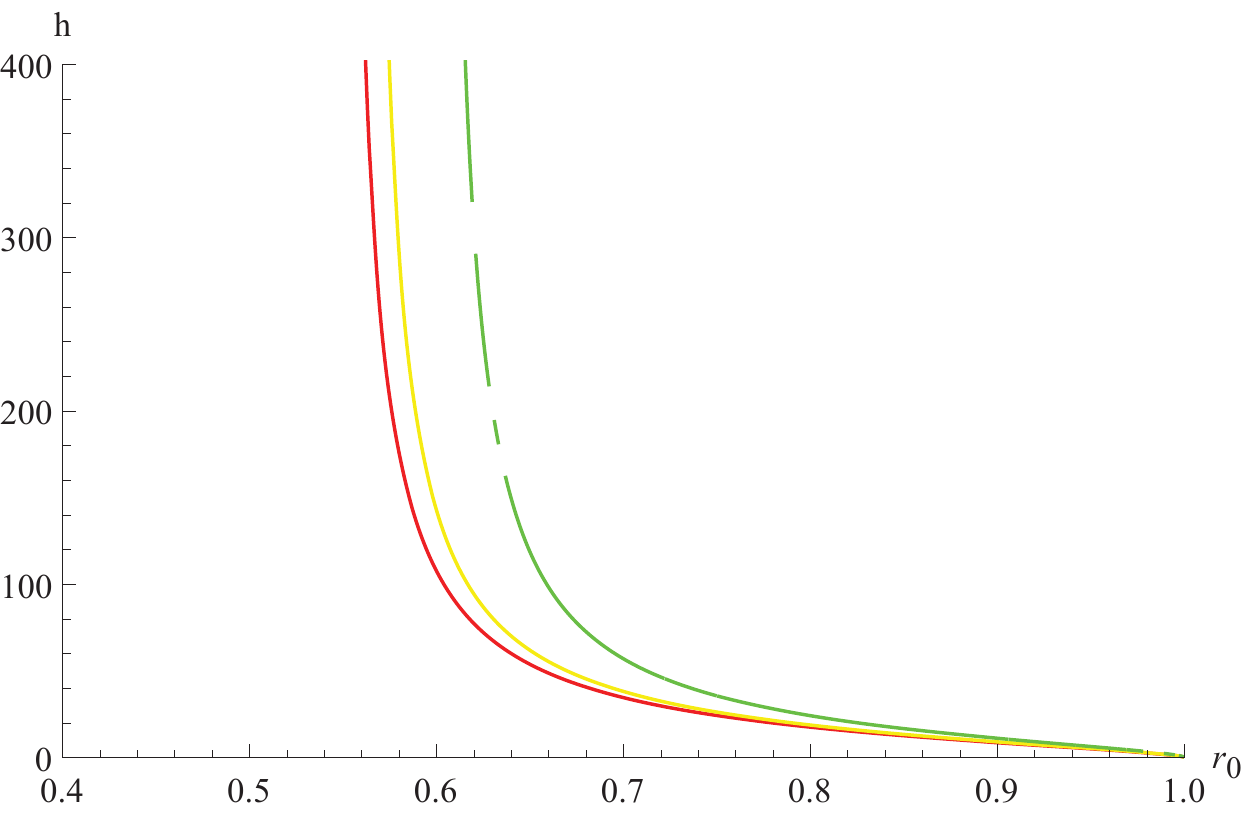}  }
\subfigure[]{
\includegraphics[trim=8.5cm 6.35cm 8.5cm 6.35cm, clip=true, scale=0.55]{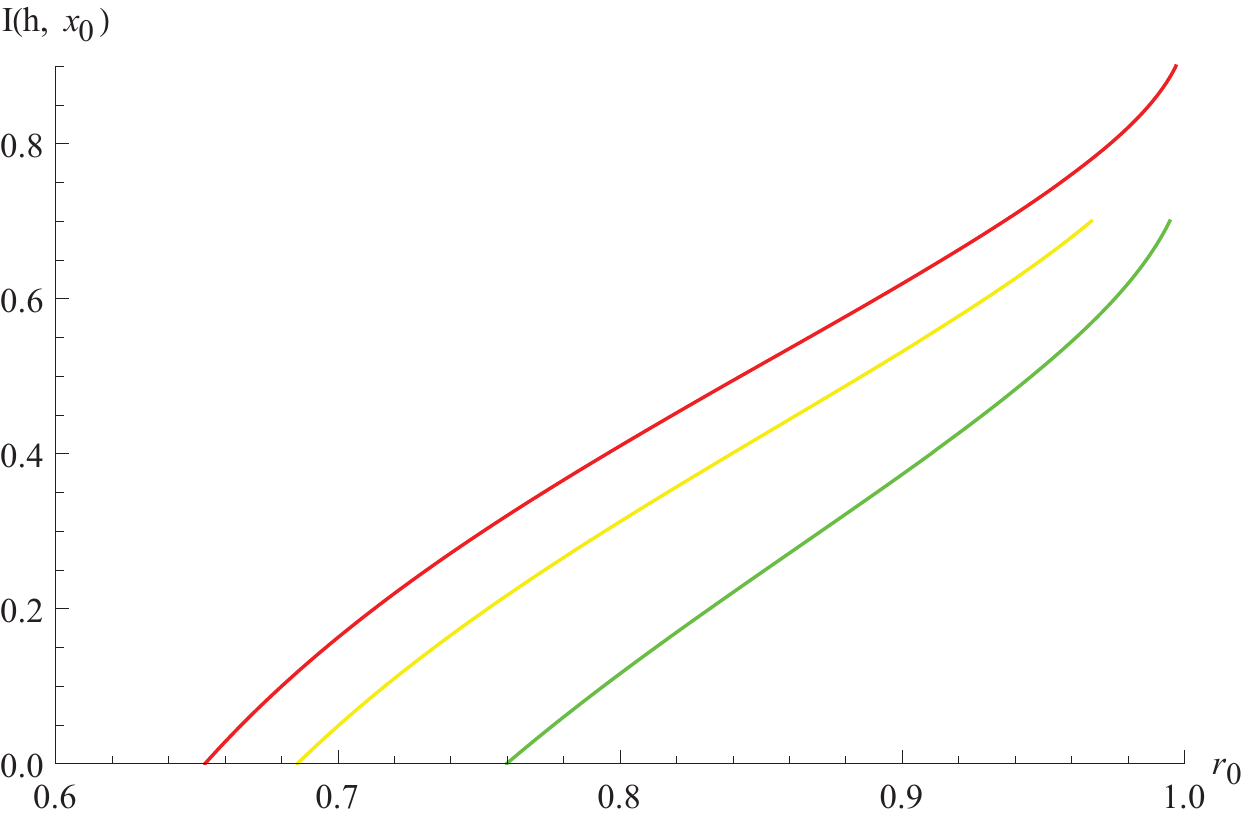}  }\\
\subfigure[]{
\includegraphics[trim=8.5cm 6.35cm 8.5cm 6.35cm, clip=true, scale=0.55]{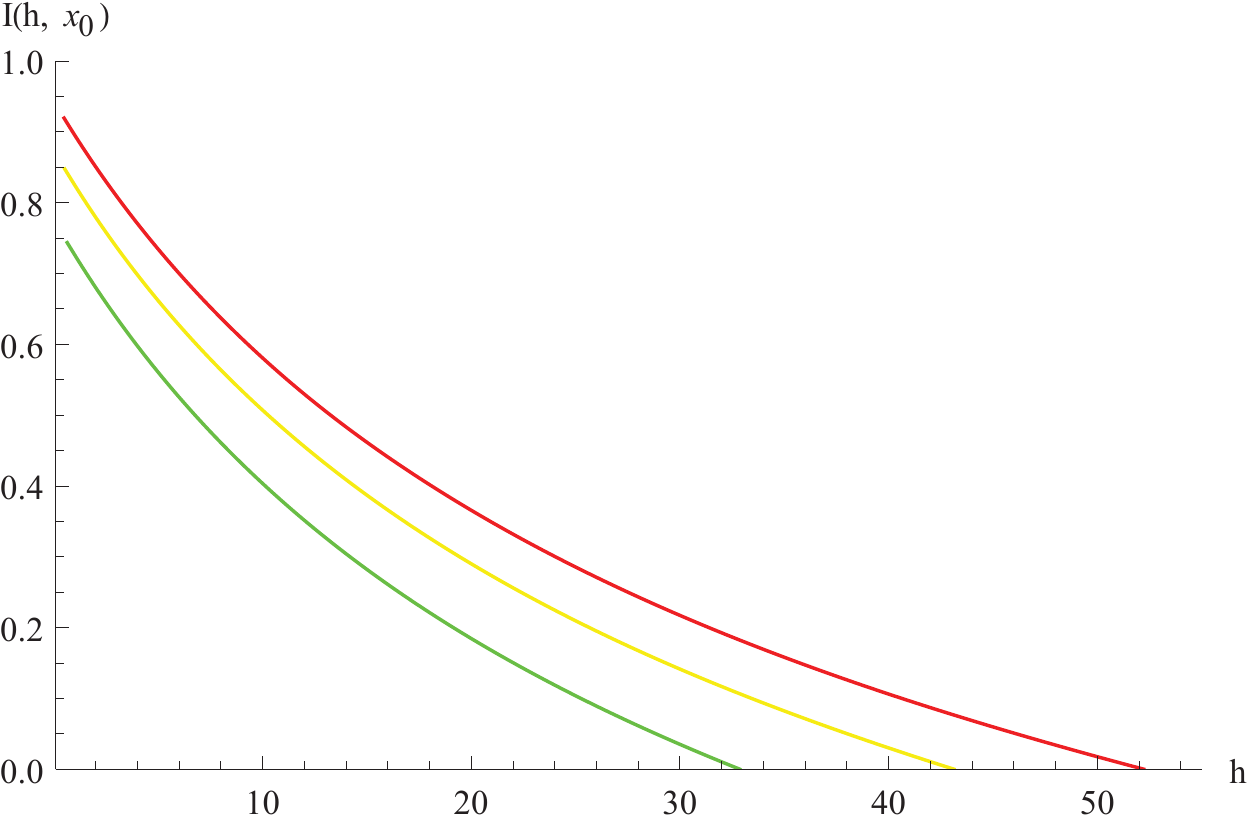}  }
 \caption{\small Panel (a): The relation between the shift $h$ and the minimal radius $r_0$ for various $m$; Panel (b): The relation between  the mutual information $I(h, x_0)$ and  $r_{0}$  for various $m$; Panel (c):   The relation between  $I(h, x_0)$ and  $h$  for different $m$. In numerics we have set $ \bar{r}=0.2$, $r_{min} = 50$, $Q=0.5, r_h=1$. The red, yellow and green lines correspond to $m=0.2, 0.3, 0.4$ respectively.} \label{fig17}
\end{figure}
Plot (a) of Fig.\ref{fig17} shows the relation between the shift $h$ and $r_0$ for different graviton masses. Just like the case in the preceding subsection, as $r_0\to r_h=1$, $h$ vanishes; Besides, there is also a critical value of $r_0$ that makes the shift $h$ diverge. The critical values of $r_0$ match  the ones from Eq.\eqref{rcrit}. For a fixed $r_0$, larger $m$ corresponds to larger $h$ according to the panel (a) of Fig.\ref{fig17};  Meanwhile we find from the panel (b) that larger $m$ corresponds to smaller mutual information, which indicates that larger inhomogeneity will disrupt the mutual information. This is consistent with the statements in the static case in the previous section.

The relationship between the mutual information and the shift $h$ is shown in the plot (c) of Fig.\ref{fig17}, which can be obtained from the rest two plots (a) and (b). Just like in the plot (c) of Fig.\ref{fig16}, the mutual information  decreases according to the shift $h$, which means the added perturbation will destroy the mutual information between the two sides of the black hole. And for a fixed value of the shift, the mutual information decreases with respect to the graviton mass as we already discussed above. There are also critical shifts $h_c$ that render the mutual information vanishing, and it is found that larger $m$ corresponds to smaller $h_c$. We have also checked other values of $m$ for these relations and found that they have  similar behaviors as in  Fig.\ref{fig17}.

\section{Conclusions and Discussions}
\label{sect:con}
In this paper we studied the holographic mutual information in the background of massive gravity. In the static case, we found that for shorter strips the near-homogeneous mutual information would increase as the temperature of the black hole grows, which is consistent with the conclusions in the pure homogeneous case studied previously. However, for a larger graviton mass, which corresponds to greater inhomogeneity in the boundary field theory, we found that the mutual information would decrease with respect to the graviton mass. For longer strips the mutual information would decrease monotonically with respect to the graviton mass, which implies that the spatial inhomogeneity has  larger influence  to the mutual information of strips with larger length. With the above results, we argued that when the system is far from the homogeneity, the spatial inhomogeneity plays a dominant role in affecting the mutual information than the temperature of the black hole.
We also investigated the effect of the charge on the mutual information and found that the mutual information decreases as the charge increases, which is independent of the width of the strip.

By adding the light-like perturbations into the bulk, we studied the dynamical mutual information in the shock wave geometry of the massive gravity. The added perturbations produce a shift on the horizon in the Kruskal coordinates. From the existing studies of the shock wave geometry, we know that the shift is  proportional to the added energy. We found that the more the added energy was, the smaller the mutual information would be, which suggests that the added perturbations would reduce the mutual information between the two sides of the black hole. We also investigated the effect of the charge and graviton mass on the critical values of the shift, where  makes the dynamical mutual information  vanish. We found that the larger the values of the charge and graviton mass are, the  smaller the critical values of the shift are, which indicates that both the charge and inhomogeneity would  reduce the mutual information.

It would be interesting to study the holographic mutual information in the background of real spatial inhomogeneity, such as adding lattice structures in the bulk \cite{Hellerman:2002qa} or for simplicity with the spatially dependent sources on the boundary \cite{Andrade:2013gsa}. The advantage of the latter is that the bulk spacetime is homogeneous and isotropic; Moreover, turning on the sources of the massless scalar fields on the boundary would provide more physical meanings of the momentum relaxation in the boundary field theory. It was shown in \cite{Andrade:2013gsa} that the parameters related to the scalar sources are similar to the graviton mass in some sectors of the massive gravity. Therefore, it would be very interesting to check whether the parameters of the scalar sources really play the same role as the graviton mass to the holographic mutual information in the massive gravity. We will leave this as our future study.

\section*{Acknowledgements}{This work is supported  by the National
Natural Science Foundation of China (Grant Nos. 11375247, 11435006, 11447601, 11405016, 11575270,  and 11675140), by a key project of CAS, China Postdoctoral Science Foundation (Grant No. 2016M590138), Natural Science Foundation of  Education Committee of Chongqing (Grant No. KJ1500530), and Basic Research Project of Science and Technology Committee of Chongqing (Grant No. cstc2016jcyja0364).
}


\end{document}